\documentclass[aps,revtex,prm,twocolumn,amsmath,superscriptaddress,showpacs,floatfix,reprint]{revtex4-2}
\usepackage{amsmath, nccmath}
\usepackage{amssymb}
\usepackage{bm}
\usepackage{braket}
\usepackage{natbib}
\usepackage{leftidx}
\usepackage{graphicx}    
\usepackage{verbatim}   
\usepackage{color}      
\usepackage{subfigure}  
\usepackage{hyperref}   
\usepackage{dcolumn}    
\usepackage{textcomp}
\usepackage{float}
\usepackage{threeparttable}
\usepackage{titlecaps}
\usepackage{multirow}
\usepackage{lipsum}
\hypersetup{
	colorlinks=true,
	citecolor=blue,
	filecolor=black,
	linkcolor=blue,
	urlcolor=blue
}
\usepackage{mathastext}
\usepackage{mathptmx}
\usepackage{enumitem}
\DeclareGraphicsExtensions{.png,.pdf,.tif}
\usepackage{pict2e}
\usepackage[dvipsnames]{xcolor}
\usepackage[normalem]{ulem}

\begin{document}

\title{Identifying open-orbit topological surface states in dual topological semimetal TaSb$_2$}


\author{Susmita Changdar}
\email{s.changdar@ifw-dresden.de}

\affiliation{Leibniz Institute for Solid State and Materials Research, IFW Dresden, 01069 Dresden, Germany}

\affiliation{Condensed Matter and Materials Physics Department, S. N. Bose National Centre for Basic Sciences, Kolkata, West Bengal-700106, India}

\author{Heike Schl\"orb}

\affiliation{Leibniz Institute for Solid State and Materials Research, IFW Dresden, 01069 Dresden, Germany}

\author{Oleksandr Suvorov}
\affiliation{Leibniz Institute for Solid State and Materials Research, IFW Dresden, 01069 Dresden, Germany}

%
\author{Dimitry Efremov}
\affiliation{Leibniz Institute for Solid State and Materials Research, IFW Dresden, 01069 Dresden, Germany}

\author{Alexander Yaresko}
\affiliation{Max Planck Institute for Solid State Research, Heisenbergstrasse 1, 70569 Stuttgart, Germany}

\author{Rui Lou}
\affiliation{Leibniz Institute for Solid State and Materials Research, IFW Dresden, 01069 Dresden, Germany}
\affiliation{Helmholtz-Zentrum Berlin für Materialien und Energie, Elektronenspeicherring BESSY II, 12489 Berlin, Germany}

\author{Alexander Fedorov}
\affiliation{Leibniz Institute for Solid State and Materials Research, IFW Dresden, 01069 Dresden, Germany}
\affiliation{Helmholtz-Zentrum Berlin für Materialien und Energie, Elektronenspeicherring BESSY II, 12489 Berlin, Germany}

\author{Bernd B\"{u}chner}
\affiliation{Leibniz Institute for Solid State and Materials Research, IFW Dresden, 01069 Dresden, Germany}

\author{Andy Thomas}

\affiliation{Leibniz Institute for Solid State and Materials Research, IFW Dresden, 01069 Dresden, Germany}

\author{Sergey Borisenko}
\email{s.borisenko@ifw-dresden.de}
\affiliation{Leibniz Institute for Solid State and Materials Research, IFW Dresden, 01069 Dresden, Germany}

\author{Setti Thirupathaiah}
\email{setti@bose.res.in}
\affiliation{Condensed Matter and Materials Physics Department, S. N. Bose National Centre for Basic Sciences, Kolkata, West Bengal-700106, India}

\date{\today}
\begin{abstract}

TaSb$_2$, a member of the transition metal dipnictide family of materials, hosts the very rare dual topological phase - weak topological insulating state and topological crystalline insulating state along different crystallographic orientations. So far, studies on the electronic structure of transition metal dipnictides have focused on their overall electronic structure and the bulk open-orbit Fermi surfaces. Using angle-resolved photoemission spectroscopy, density functional theory calculations, and transport measurements, we distinguish the intertwined bulk and surface states on the weakly topological $(20\bar{1})$ plane of TaSb$_2$. We identify multiple electron- and hole-like bulk bands, yielding a near-perfect carrier compensation. Crucially, we observe open-orbit FSs parallel to $\bar{L}$–$\bar{Y}$ direction that are entirely of surface origin. Circular-dichroism ARPES reveals $k \rightarrow -k$ spectral reversal, indicating  spin–momentum locking and the topological nature of these surface states. Consistent with this, magnetotransport measurements display weak antilocalization, establishing TaSb$_2$ as a platform for spin-polarized topological transport on a weakly topological surface.

\end{abstract}

\maketitle

\newpage
\section{Introduction}

Since the discovery of the quantum Hall effect \cite{Klitzing1980}, the interest in realizing nontrivial band topology in various materials has grown significantly. Over the years, this field has expanded from quantum Hall effect to topological insulators (TIs) \cite{Kane2005a,Kane2005b,hsieh2008,Moore2010}, Dirac \cite{Wang2012,Borisenko2014,Neupane2014} and Weyl semimetals \cite{Wan2011,xu2015,Weng2015,huang2016}, topological superconductors \cite{Fu2008,Lutchyn2010,kozii2016}, and the other classifications of topological materials \cite{Bradlyn2016,Tang2017,schindler2018}. The surface/edge states in these materials are spin–momentum locked and protected by symmetries, making them robust against disorder and promising for applications in spintronics and quantum computing \cite{Nayak2008,Stern2013}. Followed by the discovery of Weyl semimetallic phase and Fermi arcs in transtion metal pnictides (TaAs, TaP, NbAs) \cite{Weng2015,xu2015,xu2015b,xu2015c}, the transition metal dipnictides TmPn$_2$ (Tm=Ta, Nb, and Pn=P, As, Sb) attracted a lot of attention for hosting many fascinating properties including topologically non-trivial electronic states \cite{Xu2016,shao2019,Wang2019,liu2020,Regmi2023,li2022,lee2024}, giant Nerst thermopower \cite{Li2022a,Wu2025}, extremely large magnetoresistance (XMR) \cite{Li2016,Wang2016,guo2018,Ian2018,pariari2018,Lou2017, lou2022}, negative MR \cite{Li2016,Shen2016,Wang2019a}, and superconductivity \cite{li2018,Zhang2020}.

TaSb$_2$, a member of the TmPn$_2$ family, behaves as a nodal line semimetal in the absence of spin-orbit coupling (SOC). Due to band inversion, SOC opens a gap at the nodal line \cite{Xu2016, Wang2019}. Based on the $\mathbb{Z}_2$  invariants $(\nu_{0};\nu_1 \nu_2 \nu_3)$, TaSb$_2$ is a weak topological insulator (WTI) ($0;111$) \cite{Xu2016}. strong topological insulators (STIs) are classified with strong topological invariant, $\nu_0=1$,  whereas for WTIs $\nu_0=0$ with one of the invariants ($\nu_1 \nu_2 \nu_3$)  being  nonzero \cite{Fu2007,Hasan2010,Pauly2016}. Based on these weak invariants, the surface states in WTIs are only protected on certain directions. Moreover, TaSb$_2$ has monoclinic crystal structure which hosts $C_{2h}$ point group symmetry \cite{Song_2018,Wang2019,Cao2021}. $C_{2h}$ consists of mirror, rotational, and inversion symmetry. TaSb$_2$ hosts nontrivial surface states protected by rotational symmetry $C_{2(010)}$ along (010) plane, making it a topological crystalline insulator (TCI) \cite{Wang2019,Cao2021}. The coexistence of WTI and TCI states on different facets makes TaSb$_2$ particularly compelling. While similar coexisting topological phases have been observed in layered materials \cite{eschbach2017,avraham2020,Wang2022}, their topologically relevant surfaces are difficult to access. In contrast, the intrinsically three-dimensional nature of TaSb$_2$ provides an ideal platform to directly probe and manipulate surface states protected by distinct symmetries, enabling exploration of rich underlying physics as well as potential applications in low-dissipation electronics and next-generation spintronic devices \cite{brahlek2020, gilbert2021}.

In this work, we investigate the bulk and surface electronic structure of TaSb$_2$ and their impact on magnetotransport properties. Using Angle-resolved photoemission spectroscopy (ARPES) alongside first-principles calculations, we identify topological surface states coexisting with bulk bands at the Fermi level. Magnetotransport measurements reveal pronounced anisotropy arising from the anisotropic Fermi surface (FS) of TaSb$_2$ and signatures of quantum interference effects linked to these surface states. Our results establish TaSb$_2$ as a compelling platform to study the interplay between distinct topological phases and transport phenomena.

\section{Experimental details}
Single crystals of TaSb$_2$ were grown by chemical vapor transport method. In this method, initially we prepare the pollycrystalline sample of TaSb$_2$. To prepare, stoichiometric amount of Ta and Sb were thoroughly mixed together and kept in vacuum sealed quartz ampule which was then heated at 650$^o$C for 12 hours and then at 750$^o$C for 72 hours in muffle furnace. The obtained pollycrystalline TaSb$_2$ sample was then vacuum sealed in quartz ampule (10$^{-4}$ mbar) with Iodine as the transport agent (Iodine concentration: 10 gm/cc). The quartz ampule was then kept inside the gradient tube furnace for 14 days with the source and sink temperatures set at 1050$^o$C and 950$^o$C, respectively. The furnace was then cooled down to the room temperature. In this way, we obtained several single crystals of TaSb$_2$ with a typical size of $2\times3$ mm$^2$.

ARPES measurements on TaSb$_2$ were performed on the ($20\Bar{1}$) plane. The experiments were carried out at 1$^3$ ARPES end station in BESSY-II (Helmholtz Zentrum Berlin) synchrotron radiation center. Measurements were performed with angular resolution of 0.2$^o$. During the measurements the sample temperature was kept at 1.5 K, and the photon energy was varied within the range of 30-100 eV. The overall energy resolution was within 10-15 meV. Throughout the measurements the chamber pressure was kept at $9\times10^{-11}$ mBar.

 Magnetotransport measurements on TaSb$_2$ single crystals were performed in an Oxford Instruments cryostat with a variable temperature insert (VTI) and a superconducting split coil magnet, which allows to apply external magnetic fields of up to 7 T. Moreover, the cryostat is equipped with a stepper motor rotation platform that allows the sample to rotate in the magnetic field in a fixed plane. The current was applied  using a Keithley 2450 sourcemeter and both the longitudinal and transversal voltages were measured using Keithley 2182A nanovoltmeters.

\section{Computational details}

 In order to obtain the electronic band structure, calculations were performed within the framework of the DFT. The full relativistic generalized gradient approximation in the Perdew-Burke-Ernzerhof variant is employed for the exchange correlation potential, as implemented in the full potential local orbital band structure (FPLO) package \cite{Koepernik1999,fplo}. For the Brillouin zone integration, the tetrahedron method with a 12×12×12 mesh is utilized.
	 
The obtained states were employed in calculations in a semi-slab geometry, utilizing the same computational package. The Bloch wave functions were projected onto atomic-like Wannier functions, and a tight-binding model Hamiltonian was constructed. Subsequently, the tight-binding model was mapped onto a slab geometry. 

Bulk band structure calculations were performed using PY LMTO \cite{antonov2004} implementation of the Linear Muffin Tin Orbitals (LMTO) method. Spin-orbit coupling was added to the LMTO Hamiltonian at the variational step.

\section{Results and discussion}

\begin{figure}
    \centering
    \includegraphics[width=\linewidth]{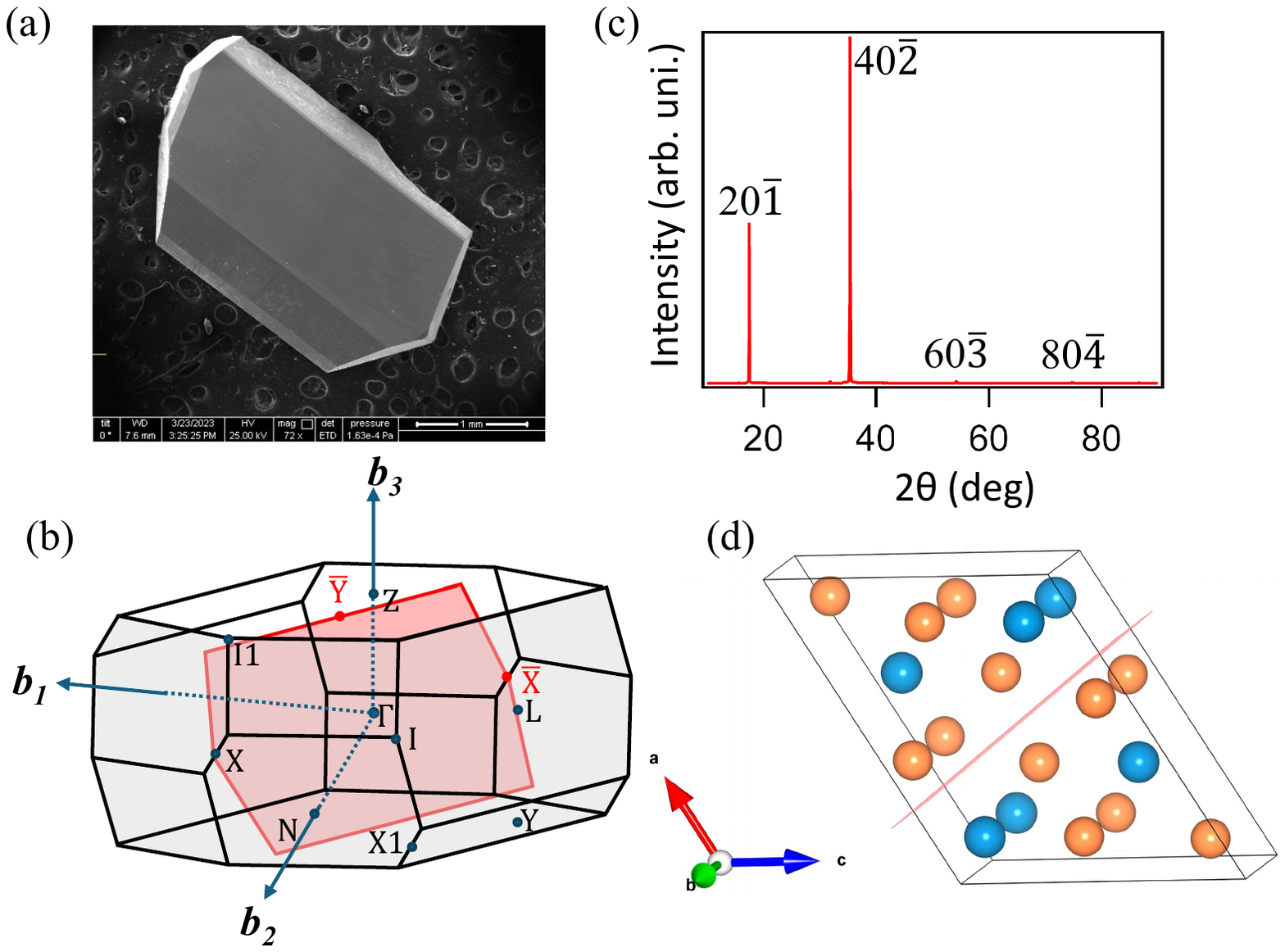}
    \caption{ Crystal structure of TaSb$_2$. (a) SEM image of the TaSb$_2$ single crystal. (b) The primitive BZ of TaSb$_2$ with ($1\Bar{1}\Bar{1}$) plane. \textit{b$_1$}, \textit{b$_2$} and \textit{b$_3$} are the reciprocal lattice vectors for the primitive unit cell. The $(1\bar{1}\bar{1})$ plane corresponds to the ($20\bar{1}$) plane for conventional unit cell~\cite{Xu2016}, which we adopt for all subsequent references. (c) The XRD data collected from ($20\Bar{1}$) plane of TaSb$_2$. (d) Schematic of the conventional unit cell of TaSb$_2$ with the ($20\Bar{1}$) plane. The blue atoms stand for Ta, and orange atoms are Sb. }
    \label{fig:1}
\end{figure}

\begin{figure*}
    \centering
    \includegraphics[clip = true, width=\linewidth, trim = 0 2cm 0 1cm ]{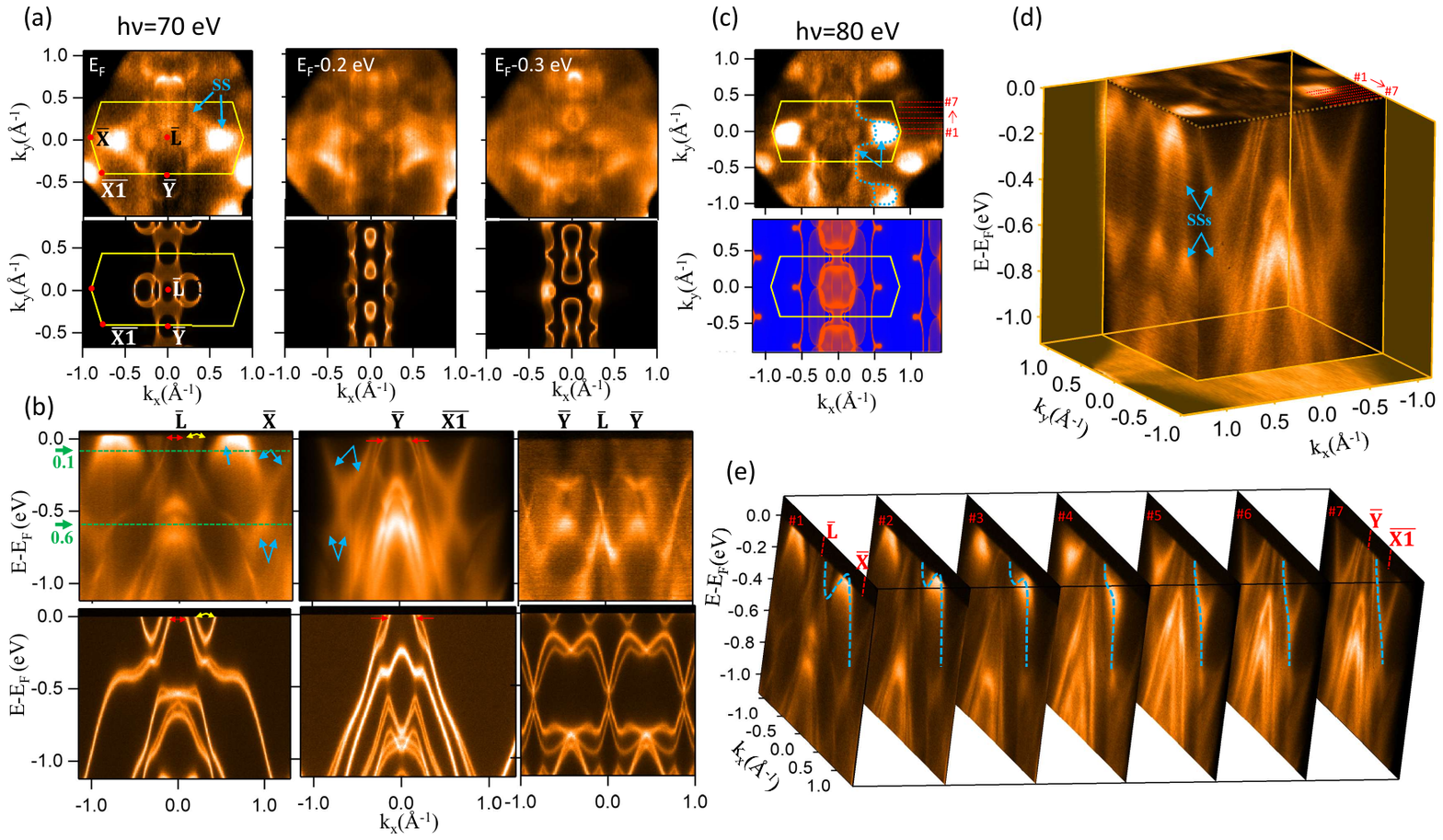}
    \caption{In-plane electronic structure of TaSb$_2$. (a) Top: FS map (left) and constant energy contours at 0.2 eV (middle) and 0.3 eV (right) below the Fermi level from ARPES ($h\nu$ = 70 eV). Bottom: corresponding bulk band calculations. (b) Top: EDMs along $\Bar{L}-\Bar{X}$ (left), $\Bar{Y}-\Bar{X1}$ (middle), and $\Bar{L}-\Bar{Y}$ (right). Bottom: corresponding bulk calculations. Blue arrows point the surface states. Green lines (top-left panel) show the momentum distribution curve (MDC) used in the out-of-plane contours in Fig.~\ref{fig:3}(a).  (c) FS map from ARPES (top) compared with surface state calculations (bottom). (d) 3D ARPES spectra showing the in-plane dispersion of the marked surface state. (e) Evolution of the band structure from $\Bar{L}-\Bar{X}$ toward $\Bar{Y}-\Bar{X1}$. Red dotted lines in (c)–(d) mark the positions of the EDM cuts.}
    \label{fig:2}
\end{figure*}

TaSb$_2$ crystallizes in a monoclinic structure with space group \textit{C}12/\textit{m}1. Scanning electron microscopy (SEM) image of a typical single crystal is shown in Fig.~\ref{fig:1}(a), displaying its well-faceted morphology. To define the ARPES measurement geometry, we present the three-dimensional Brillouin zone (BZ) of the primitive unit cell, along with its projection onto the $(1\bar{1}\bar{1})$ plane (equivalent to ($20\Bar{1}$) plane for conventional unit cell) in Fig.~\ref{fig:1}(b). The X-ray diffraction (XRD) data, shown in Fig.~\ref{fig:1}(c), confirm that the cleaved surface is indeed the ($20\bar{1}$) plane. Furthermore, we demonstrate the conventional unit cell of TaSb$_2$, highlighting the arrangement of Ta and Sb atoms along with the ($20\bar{1}$) cleavage plane. This visualization helps to establish the correspondence between real-space orientation and the measured momentum space, providing a good understanding of the structural framework for interpreting the surface electronic structure discussed below.

Figure \ref{fig:2}(a) displays the in-plane FS map, constant energy contours at $E_F - 0.2$ eV, and $E_F - 0.3$ eV, measured with $p$-polarized light at a photon energy of $h\nu = 70$ eV. The corresponding calculated bulk band structures shown in the lower panels reveal excellent agreement with experimental intensity distributions near $\Bar{L}$ and $\Bar{Y}$, confirming the bulk origin of states in these high symmetry points. A 50 meV shift towards higher binding energy was applied to all of the calculated bands to align them with experimental Fermi level positions. We focus on the FS centered at $\Bar{L}$—located at $k_z = \pi$—rather than at $\Bar{\Gamma}$, where no spectral weight is present. Focusing on the ARPES spectra in Fig. \ref{fig:2}(a), we observe a ripple-like, open-orbit structure near $\Bar{X}$. This feature is absent in bulk band calculations but emerges clearly in the semi-infinite slab calculations [Fig. \ref{fig:2}(c) bottom panel], confirming its surface origin.
\begin{figure*}[!ht]
    \centering
    \includegraphics[clip = true, width=0.9\linewidth, trim = 0 3.3cm 0 3cm ]{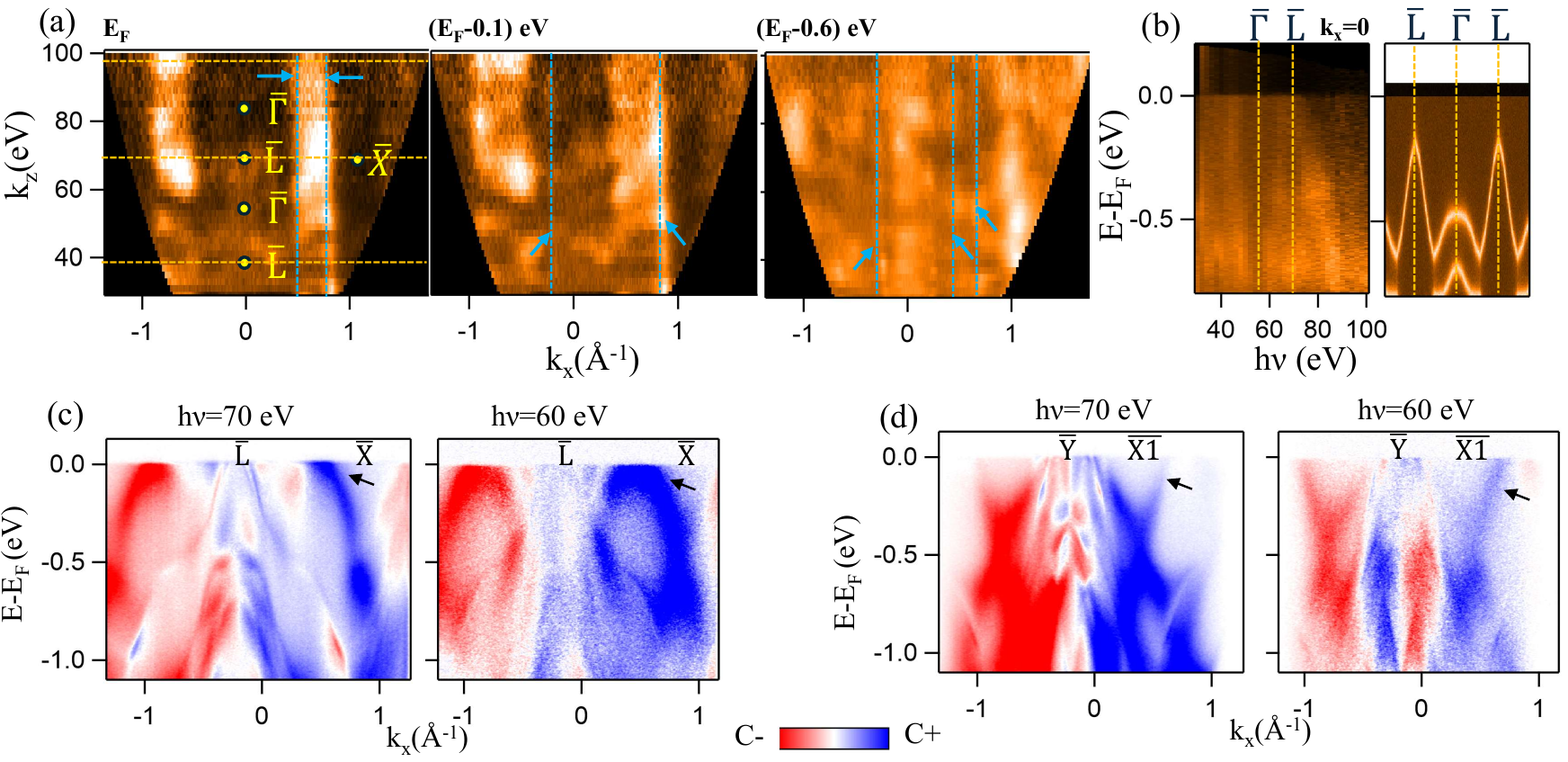}
    \caption{Out-of-plane electronic structure and polarization-dependent ARPES measurements. (a) Out-of-plane FS, constant energy contours acquired at 0.1 eV, and 0.6 eV below the Fermi level respectively. The blue dotted lines indicate the surface states exhibiting negligible k$_z$ warping. (b)  Left panel: Out-of-plane EDM obtained at k$_x$=0$\AA^{-1}$. Right panel: Corresponding bulk calculation. (c) EDM along $\Bar{L}-\Bar{X}$ acquired by subtracting the spectral intensity obtained from right circularly polarized light (C+) and left circularly polarized light (C-) with photon energies of 70 eV and 60 eV. (d) represents the same for the EDM taken along $\Bar{Y}-\Bar{X1}$. }
    \label{fig:3}
\end{figure*}
Figure \ref{fig:2}(b) presents energy–momentum dispersions (EDMs) along high-symmetry directions, with corresponding DFT results shown below each panel. These comparisons enable a clear distinction between bulk and surface-derived states (the latter highlighted by blue arrows). Along $\Bar{L}-\Bar{Y}$, the experimental bands align well with bulk calculations, reinforcing their bulk origin. Several dispersive bands crossing the Fermi level give rise to both electron and hole pockets, consistent with a semimetallic electronic structure. Along $\Bar{L}-\Bar{X}$, both electron-like (marked with yellow arrows) and hole-like dispersions (marked with red arrows) are observed. In Fig. \ref{fig:2}(c), we compare the FS map with DFT slab calculations. The momentum-space trajectory of the surface band is highlighted with blue dashed lines and is generally consistent with the surface bands observed in the calculated FS map, except that the surface state shows less dispersion along $k_x$ compared to the ARPES results. To visualize the surface-state dispersion across the Brillouin zone, Fig. \ref{fig:2}(d) shows a 3D ARPES intensity map. Both ARPES and slab calculations in Fig. \ref{fig:2}(c) reveal that the surface band crosses the Fermi level once along $\Bar{Y}-\Bar{X1}$, but twice along $\Bar{L}-\Bar{X}$. To elucidate this, EDMs along paths \#1–\#7 [Fig. \ref{fig:2}(e)] are plotted. The dotted lines are a guide for the eye to show how the band dispersion evolves from an electron-like dispersion near $\Bar{Y}-\Bar{X1}$ to a mixed electron–hole-like character near $\Bar{L}-\Bar{X}$.

To showcase that the bands labeled as surface states in Fig. \ref{fig:2} have minimal $k_z$ warping, we plotted the out-of-plane FS map, and the constant energy contours at $E_F$-0.1 eV and $E_F$-0.6 eV [Fig. \ref{fig:3}(a)]. The contours showcase minimal $k_z$ warping for the previously marked surface bands in the EDM along $\Bar{L}-\Bar{X}$ [Fig. \ref{fig:2}(b) top-left panel], therefore reinforcing our assumption that they are $k_z$-independent surface states. Next we present the out-of-plane EDMs. EDM taken at $k_x=k_y= 0 \AA^{-1}$ show that the dispersive band touches the Fermi level at 68 eV and at around 40 eV. Since, band dispersion along $\Bar{\Gamma}-\Bar{L}$ only crosses the Fermi level at $\Bar{L}$, with no bands near $\Bar{\Gamma}$ [Fig. \ref{fig:3}(b)], we mark the $h\nu=68$ eV as a $\Bar{L}$ point, and $h\nu=54$ eV as $\Bar{\Gamma}$ point. The high symmetry points in the contour maps in Fig.\ref{fig:3}(a) are mainly determined from here. The consideration is also quite in agreement with the our consideration of the high symmetry points in Fig. \ref{fig:2}. 

The effect of dichroism on the electronic structure of TaSb$_2$ is presented in Fig.\ref{fig:3} (c)-(d). By subtracting the spectral intensity of the EDMs collected with right and left cirular polarized light, we have obtained the EDMs plotted in Fig. \ref{fig:3} (c)-(d). The extracted EDMs along $\Bar{L}-\Bar{X}$ collected with $h\nu=70$ eV and 60 eV are plotted in Fig. \ref{fig:3}(c). Here, the EDMs collected with both photon energies show that spectral intensity for the surface state switches between $k$ and $-k$. The similar switching of spectral intensity for the surface band is also present for the extracted EDMs along $\Bar{Y}-\Bar{X1}$ [Fig. \ref{fig:3}(d)]. Usually, helicity of the circularly polarized light couples with the spin degree of freedom of the electrons in the systems with spin-orbit coupling \cite{Park2011,Park2012,Wang2013}.

\begin{figure*}
    \centering
    
    \includegraphics[clip = true, width=0.87\linewidth, trim = 1.5cm 0.5cm 1.5cm 0.1cm ]{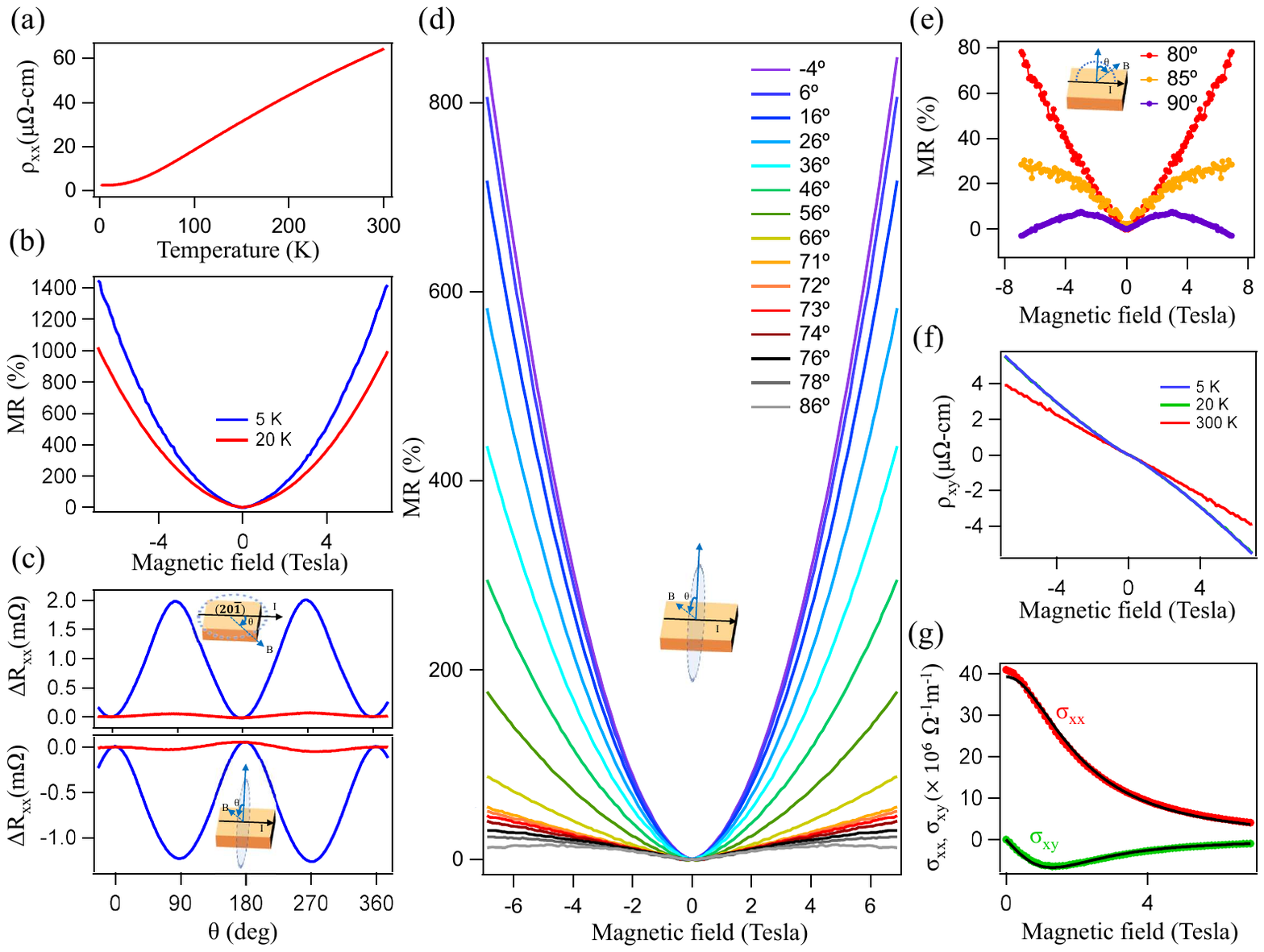}
        \caption{(a) Temperature dependence of the longitudinal resistivity, $\rho_{xx}(T)$. (b) MR as a function of magnetic field at 5 K and 20 K. (c) Anisotropic MR [$\Delta R_{xx}=R_{xx}(\theta)-R_{xx}$ (0 deg)] measured at 6 T as a function of field angle $\theta$ for in-plane (top panel) and out-of-plane (bottom panel) rotations at 5 K and 300 K. The measurement geometries are demonstrated in the schematics. MR ($B,\theta$) for field rotation in a plane oriented (d) perpendicular and (e) parallel to the current direction ($b$ axis). (f) Hall resistivity, $\rho_{xy}$, versus magnetic field at three representative temperatures. (g) Longitudinal conductivity $\sigma_{xx}$ and Hall conductivity $\sigma_{xy}$ as a function of magnetic field at 5 K, with solid lines representing the two-band model fitting.}
    \label{fig:4}
\end{figure*}

We also performed comprehensive magnetotransport measurements to directly investigate the effect of both surface and bulk states on the transport properties of TaSb$_2$. The semi-metallic nature of the sample reflects in the resistivity $\rho_{xx}(T)$ data from $(20\bar{1})$ plane. We observed large positive magnetoresistance ($\text{MR} = [\rho_{xx}(B) - \rho_{xx}(0)]/\rho_{xx}(0)$), reaching $1450\%$ at 5 K and 7 T with an almost parabolic field dependence, which is suppressed upon increasing temperature. Consistent with the low-symmetry monoclinic crystal structure \cite{pariari2018,Singha2018} and a complex FS \cite{Sun2022}, the MR is highly anisotropic [Fig.~\ref{fig:4}(c)]. We observe two-fold rotational symmetry with a minimum for the I$||$B$||$b-axis for in-plane rotation of the magnetic field. This is due to the effect of Lorentz force on the pockets maximizing when the field is perpendicular to the current direction. However, in Fig. \ref{fig:4}(c) bottom panel, the field always remains perpendicular to the current direction (I$||$b-axis) and the maximum MR occurs when B is perpendicular to both the sample surface and the current direction ($\theta$ =0$^{\circ}$). Here, $\mathrm{MR}(B,\theta)$ reveals a clear deviation from parabolic field dependence with increasing $\theta$ [Fig.~\ref{fig:4}(d)]. The deviation becomes prominent above $\theta=73^{\circ}$. As the direction of the applied magnetic field changes, the effective cross section of the highly anisotropic electron- and hole-type Fermi pockets perpendicular to the field is modified, which in turn alters the transport contributions of electron and hole carriers in TaSb$_2$. A similar sensitivity of the MR$(B)$ curvature to effective carrier imbalance was demonstrated through simulations in the related compound ZrP$_2$ \cite{Bannies2021}, where deviations from perfect electron--hole compensation led to a comparable deviation from the parabolic MR(B) curve.

For the out-of-plane field rotation with the rotation plane parallel to the current [Fig.~\ref{fig:4}(e)], the longitudinal magnetoresistance (LMR, $I \parallel B$ at $\theta = 90^\circ$) exhibits an initial increase up to $\sim 3$ T before transitioning into a negative slope at higher fields. This low-field cusp is a signature of weak antilocalization (WAL), previously observed in isostructural TaSb$_2$ \cite{Li2016} and TaAs$_2$ nanowires \cite{Roy2025}, where it was attributed to the topological surface states. Therefore, the topological surface states we have identified by ARPES, dominate the low-field transport near the in-plane orientation. At higher fields, the onset of negative LMR indicates crossover to a transport regime influenced by the chiral anomaly. Similar transitions from WAL-induced positive MR to chiral-anomaly-driven negative MR have been reported in several topological materials, including TaAs \cite{Huang2015}, MoGe$_2$ \cite{Huang2021}, and PtSe$_2$ \cite{Li2018a}. The extreme angular sensitivity of the effect, with the negative MR component disappearing when the magnetic field is tilted by only $5^\circ$ away from the longitudinal configuration ($\theta = 85^\circ$), is consistent with a transport mechanism that depends on a finite $\mathbf{E}\cdot\mathbf{B}$ term, which is required for the suppression of backscattering between Weyl/Dirac nodes.

Moreover, the Hall measurements [Fig.~\ref{fig:4}(f)] confirmed the multi-band nature and overall electron-dominated transport. A simultaneous two-band fitting of the conductivity tensors $\sigma_{xx}$ and $\sigma_{xy}$ using the standard formulas $\sigma_{xx}=\sum_{i=1}^2\frac{e n_i \mu_i}{1+\mu_i^2 B^2}$ and $\sigma_{xy}=\sum_{i=1}^2 S_i \frac{e n_i \mu_i^2 B}{1+\mu_i^2 B^2}$ (where $S_i = \pm 1$ for holes/electrons) yielded a near-perfect electron-hole compensation at 5 K, with electron carrier density $n_e = 2.78(1) \times 10^{20}$ cm$^{-3}$ and mobility $\mu_e = 5.66(3) \times 10^3$ cm$^2$/Vs, slightly exceeding the hole values ($n_h = 2.56(2) \times 10^{20}$ cm$^{-3}$, $\mu_h = 3.45(3) \times 10^3$ cm$^2$/Vs). This near perfect electron-hole compensation agrees with the  ARPES results where we see both electron and hole-like bands, and is a strong contributor to the large MR \cite{Zeng2016} we observe. Overall, these transport results demonstrate that both topological surface states and the bulk electronic structure play crucial roles in determining the low-temperature electrical properties of TaSb$_2$.

Although we were unable to directly investigate the interplay between the surface states of the TCI and weak TI phases in TaSb$_2$, as the sample could not be cleaved along the (010) plane, future studies in this direction may provide valuable insights. Though the electronic and transport properties of TaSb$_2$ \cite{Li2016,pariari2018,lee2024} have been previously investigated, in this work, for the first time, we successfully isolated the topological surface states from the bulk, and their impact in transport. We have realized that, unlike the bulk open-orbit FSs reported in other TmPn$_2$ compounds \cite{Lou2017, lou2022}, the open FSs in the $(20\bar{1})$ plane of TaSb$_2$ originate from topologically protected surface states, which gives rise to the WAL in MR. These observations mark a significant
step toward understanding the electronic structure and its
influence on the transport properties of TaSb$_2$.

\section{Conclusions}
In summary, our combined ARPES, DFT, and magnetotransport studies provide a comprehensive picture of the electronic structure of TaSb$_2$ along the weakly topological ($20\bar{1}$) plane. Comparison with DFT calculations allows us to distinguish surface states from bulk bands. Multiple electron- and hole-like bulk bands cross the Fermi level, leading to near-perfect carrier compensation that gives rise to the large MR. Distinct topological surface states appear along $\bar{L}-\bar{X}$ and $\bar{Y}-\bar{X}_1$. These surface states are open parallel to $\bar{L}-\bar{Y}$ that arise from weak topological symmetry protection. Dichroic ARPES indicates the spin-polarized nature of these open surface states, suggesting spin–momentum locking. Consistent with these observations, magnetotransport measurements show WAL at low magnetic field range. These findings provide important insight into the interplay between bulk and topological surface states and their combined impact on the magnetotransport properties of TaSb$_2$.

\section*{Acknowledgements}
We acknowledge beamtime at a 1$^3$-ARPES end-station at BESSY-II. The work at IFW and HZB was supported by the Deutsche Forschungsgemeinschaft under Grant SFB 1143 (project C04) and the Würzburg-Dresden Cluster of Excellence on Complexity and Topology in Quantum Matter – ct.qmat (EXC 2147, project ID 390858490). This work was performed during the research visit in Leibniz Institute for Solid state and Materials Research, Dresden as a part of collaboration between  Leibniz Institute for Solid state and Materials Research, Dresden and S. N. Bose National Centre for Basic Science, Kolkata. This research has made use of the Technical Research Centre (TRC) instrument facilities of the S. N. Bose National Centre for Basic Sciences, established under the TRC project of the Department of Science and Technology (DST), Govt. of India. S.C. acknowledges the support by BMBF funding through project 01DK240008  (GU-QuMat).  S.T. thanks the financial support provided by UGC-DAE CSR through grant no. CRS/2021-22/01/373. O.S. and B.B. acknowledge the support of BMBF through project “Instant micro-ARPES for in-operando tuning of material and device properties”. S.T. thanks the Science and Engineering Research Board (SERB), Department of Science and Technology (DST), India, for the financial support through grant no. SRG/2020/000393. We thank Ulrike Nitzsche for technical assistance. D.V.E. acknowledge DFG for financial support (grants 529677299, 449494427). 
	
\bibliography{TaSb2}

\end{document}